%% file: CSQCD_tex-5_Iosilevskiy.tex
\documentclass[12pt]{article}
\usepackage{epsfig}
\usepackage{graphicx}
\usepackage{subfigure}
\usepackage{sidecap}
\usepackage{amsmath}
\usepackage{diagrams}
\usepackage{lmodern}

\input econfmacros-csqcd4.tex
\def\Title#1{\begin{center} {\Large {\bf #1} } \end{center}}

\begin{document}

\Title{Entropic and enthalpic phase transitions in high energy density nuclear matter}

\bigskip\bigskip


\begin{raggedright}

{\it
Igor Iosilevskiy$^{1,2}$

\bigskip
$^{1}$
Joint Institute for High Temperatures (Russian Academy of Sciences),
Izhorskaya Str. 13/2, 125412 Moscow, Russia\\
\bigskip
$^{2}$
Moscow Institute of Physics and Technology (State Research University),
Dolgoprudny, 141700, Moscow Region, Russia\\
}

\end{raggedright}

\section{Abstract}

Features of Gas-Liquid (GL) and Quark-Hadron (QH) phase transitions (PT) in dense nuclear matter are under discussion in comparison with their terrestrial counterparts, e.g. so-called "plasma" PT in shock-compressed hydrogen, nitrogen etc. 
Both, GLPT and QHPT, when being represented in widely accepted temperature -- baryonic chemical potential plane, are often considered as similar, i.e. amenable to one-to-one mapping by simple scaling. 
It is argued that this impression is illusive and that GLPT and QHPT belong to different classes: GLPT is typical enthalpic PT 
(Van-der-Waals-like) while QHPT ("deconfinement-driven") is typical entropic PT. 
Subdivision of 1st-order fluid-fluid phase transitions into enthalpic and entropic subclasses was proposed in [arXiv:1403.8053]. 
Properties of enthalpic and entropic PTs differ significantly. Entropic PT is always internal part of more general and extended thermodynamic anomaly -- domains with abnormal (negative) sign for the set of (usually positive) second derivatives of thermodynamic potential, e.g. Gruneizen and thermal expansion and thermal pressure coefficients etc. 
Negative sign of these derivatives lead to violation of standard behavior and relative order for many iso-lines in $P$--$V$ plane, 
e.g. isotherms, isentropes, shock adiabats etc. 
Entropic PTs have more complicated topology of stable and metastable areas within its two-phase region in comparison with conventional enthalpic PTs. 
In particular, new additional metastable region, bounded by new additional spinodal, appears in the case of entropic PT. 
All the features of entropic PTs and accompanying abnormal thermodynamics region have transparent geometrical interpretation 
-- multi-layered structure of thermodynamic surfaces for temperature, entropy and internal energy as a pressure--volume functions, 
e.g. $T(P,V)$, $S(P,V)$ and $U(P,V)$.

\section{Introduction}

Phase transition (PT) is universal phenomena in many terrestrial and astrophysical applications. There are very many variants of hypothetical PTs in ultra-high energy and density matter in interiors of neutron stars (so-called hybrid or quark-hadron stars)~\cite{HPE-07}, in core-collapse supernovae explosions and in products of relativistic ions collisions in modern super-colliders (LHC, RHIC, FAIR, NICA etc.). Two hypothetical 1$^\mathrm{st}$-order phase transitions are the most widely discussed in study of high energy density matter ($\rho \sim 10^{14}$ g/cc): (\textit{\textbf{i}}) -- gas-liquid-like phase transition (GLPT) in ultra-dense nuclear matter: i.e. in equilibrium (Coulombless) ensemble of protons, neutrons and their bound clusters $\{p, n, N(A,Z)\}$ at $T \le 20$ MeV, and (\textbf{\textit{ii}}) -- quark-hadron (deconfinement) phase transition (QHPT) at $T \le 200$ MeV. (see e.g.~\cite{Fort-09, Fort-11}). Both, GLPT and QHPT, when being represented in widely accepted $T$--$\mu_B$ plane ($\mu_B$ -- baryonic chemical potential) are often considered as similar, i.e. amenable to one-to-one correspondence with possible transformation into each other by simple scaling (see e.g. figures~1 and~12 in~\cite{HDSI-13}). The main statement of present paper is that this impression is illusive and that GLPT and QHPT belong to different classes:  GLPT is typical \textit{enthalpic} (VdW-like) PT, while ``deconfinement-driven'' QHPT is typical \textit{entropic} PT (see~\cite{ILI-S-H-PT-13} and \cite{NUFRA-13}) like hypothetical ionization- and dissociation-driven phase transitions in shock-compressed dense hydrogen, nitrogen etc. in megabar pressure range (see e.g.~\cite{IoSta-00}).

It should be noted that the term "enthalpic" PT is not accepted and not used presently. As for the term "entropy-driven" PT, it is used already in application to rather delicate structural PTs (e.g.~\cite{Onsager, Frenkel-99, Zilman-03} etc). In present paper the two terms, entropic and enthalpic PTs, are offered as general and universal ones for wide number of applications (e.g.~ \cite{IoSta-00}). Fundamental difference of entropic and enthalpic PTs, defined in this way, are discussed and illustrated below.

\section{Comparison of GLPT and QHPT in density--temperature plane}

\begin{figure}[t]
\subfigure[Gas-liquid phase transition.]{
\includegraphics[width=0.47\columnwidth]{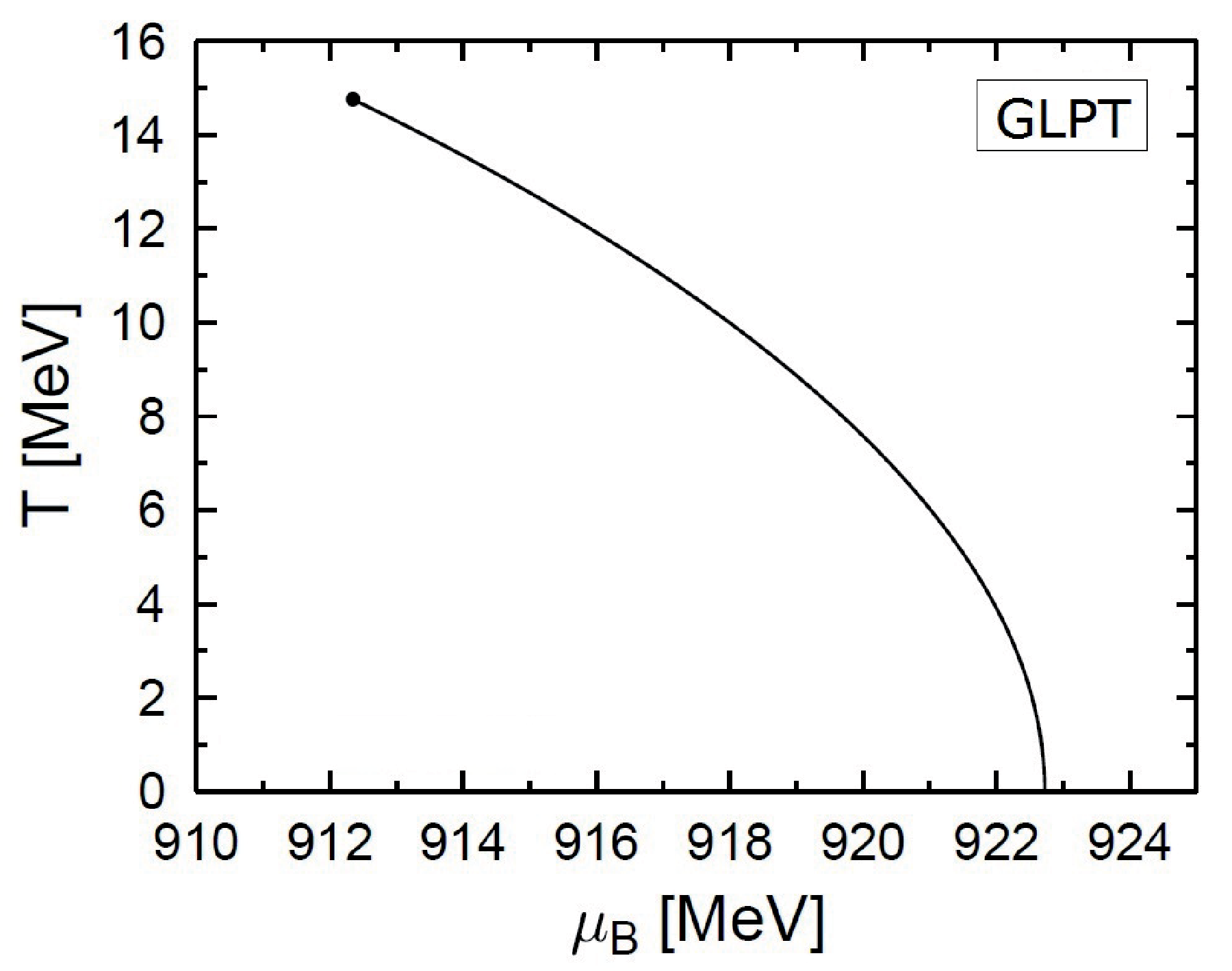}
\label{tm-glpt}}
\hfill
\subfigure[Quark-hadron phase transition.]{
\includegraphics[width=0.49\columnwidth]{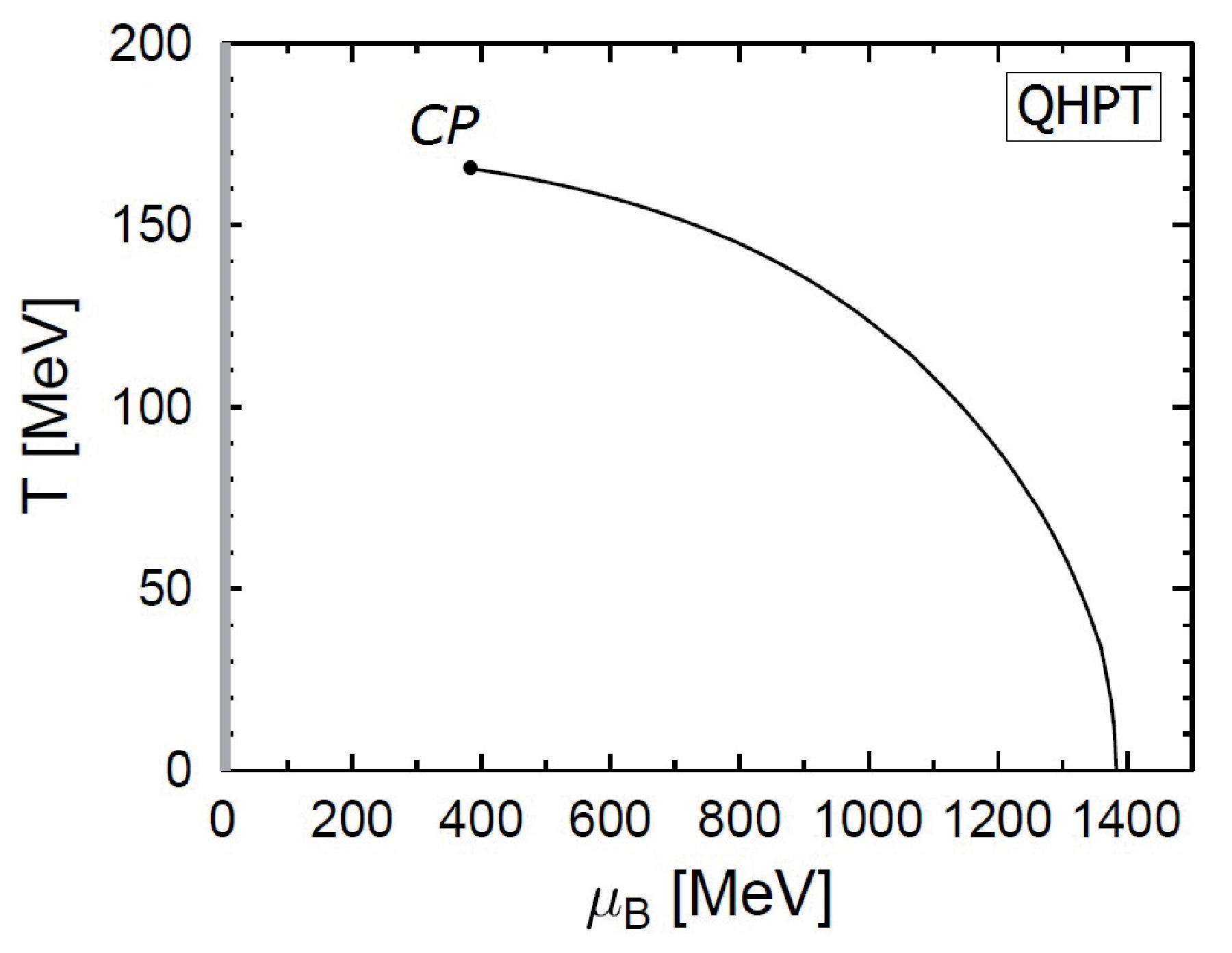}
\label{tm-qhpt}}
\caption{Visible equivalence of Gas-liquid (GLPT) and Quark-hadron (QHPT) phase transitions in symmetric nuclear matter in temperature -- baryon chemical potential plane. Figure from~\cite{HDSI-13}.
(GLPT -- FSUGold model in ensemble $\{p,n,N(A,Z)\}$, QHPT -- SU(3) model)}
\end{figure}

GLPT and QHPT look as similar in $T$--$\mu_B$ plane (figures~\ref{tm-glpt} and \ref{tm-qhpt}). It should be noted that unfortunately this type of representation is not revealing for PT analysis. Fundamental difference between GLPT and QHPT could be more evidently demonstrated in other variants of phase diagram widely used in electromagnetic plasmas community (see e.g.~\cite{IoSta-00}). First one is density--temperature $(T$--$\rho)$ diagram. Two these phase transitions (GLPT and QHPT) are sometimes considered in $T$--$\rho$ plane as quantitatively, not qualitatively different in their schematic comparison (see e.g. fig.~2 in~\cite{Rand-12} and slide 2 in~\cite{Rand-CPOD}). Numerical calculations of phase boundaries for GLPT and QHPT (see fig.~3 and~14 in~\cite{HDSI-13}) demonstrate significant difference in structure of these two boundaries (fig.~2a below~\cite{HD-Priv}).

\begin{figure}[t]
\subfigure[Liquid-gas and quark-hadron phase transitions]{
\includegraphics[width=0.49\columnwidth]{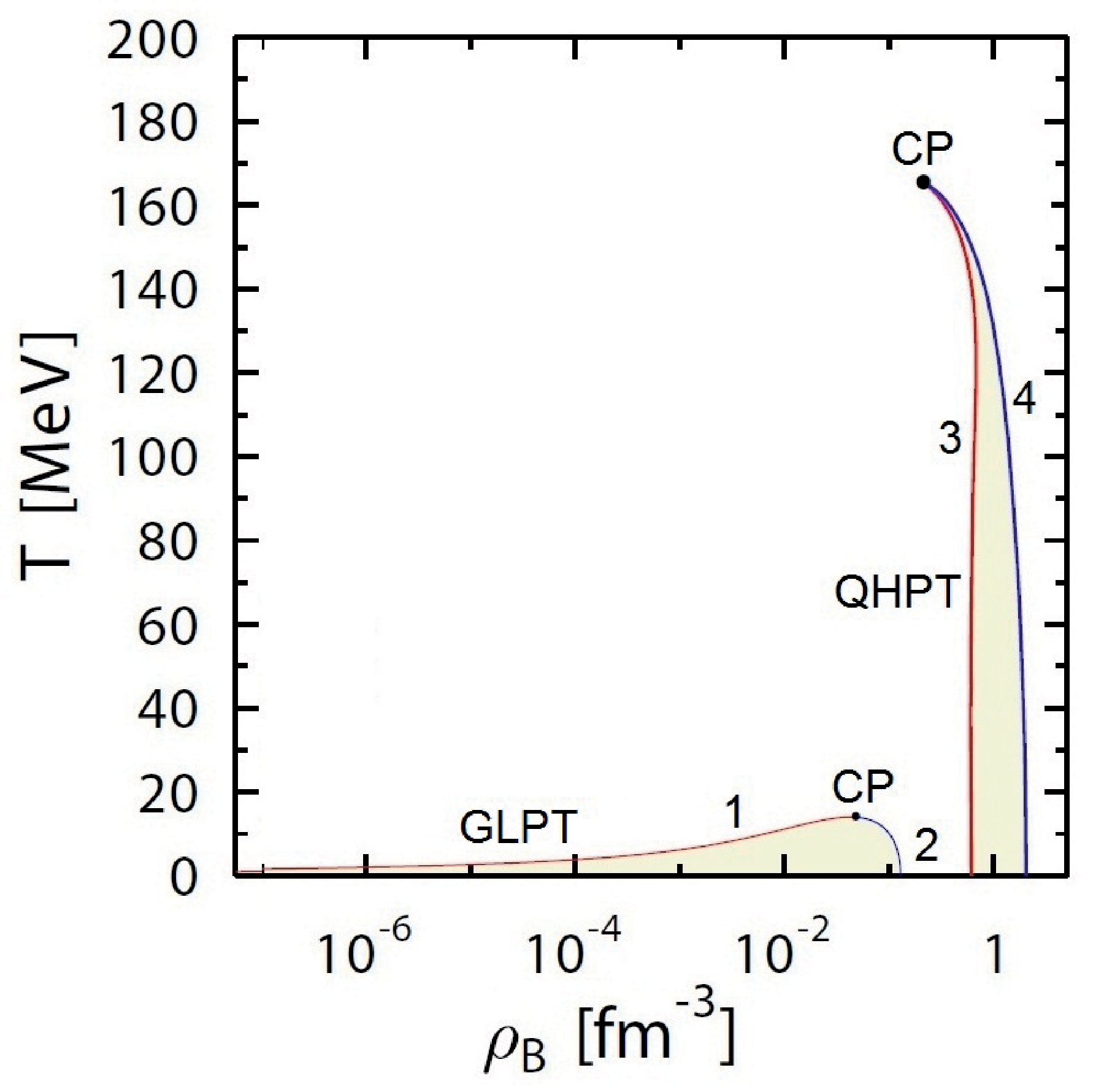}
\label{tr-1.png}}
\hfill
\subfigure[Gas-liquid and plasma phase transitions]{
\includegraphics[width=0.46\columnwidth]{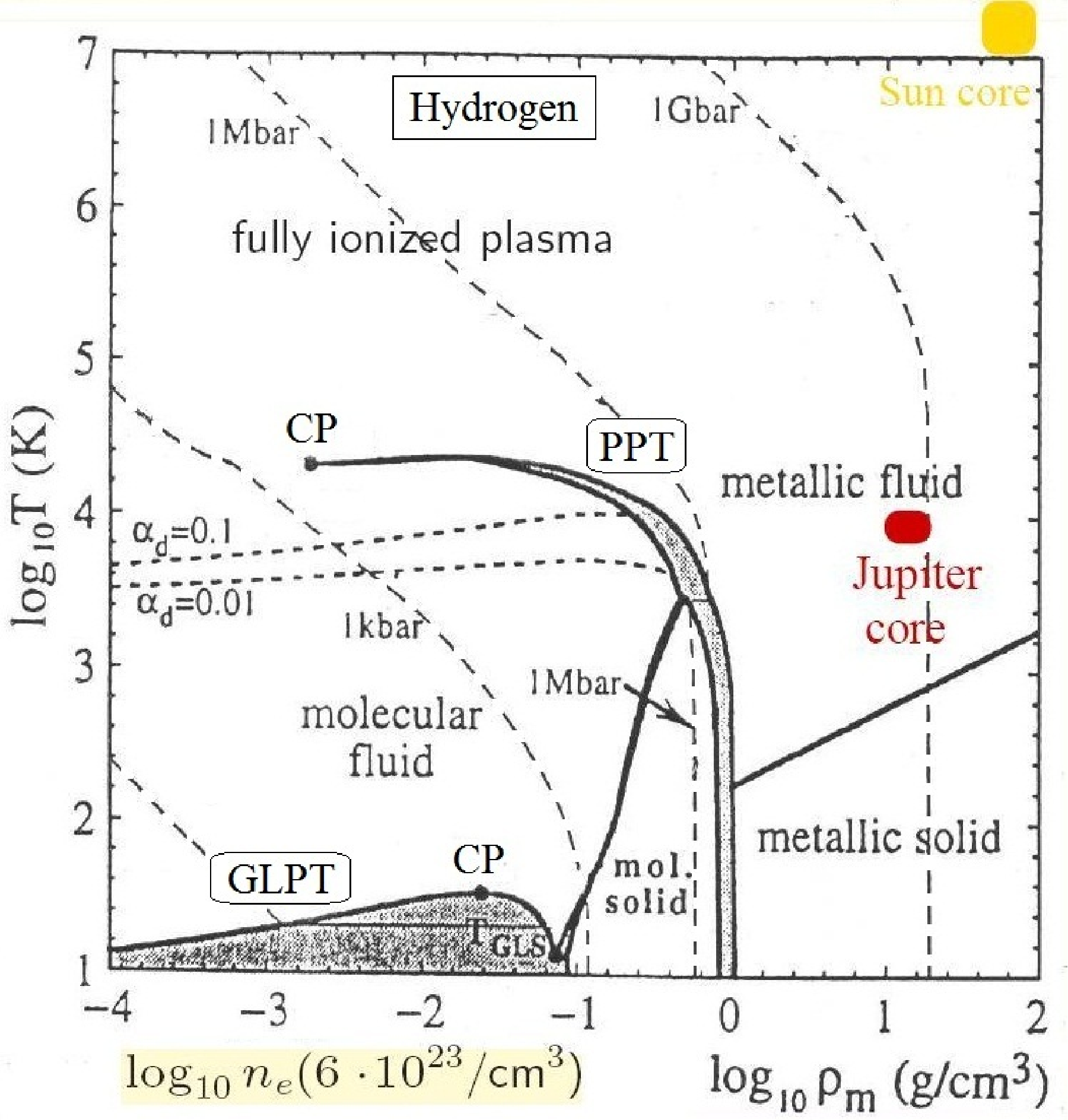}
\label{tr-h.png}}
\caption{(a): Gas-liquid and quark-hadron phase transitions (GLPT vs. QHPT) in symmetric nuclear matter~\cite{HDSI-13, HD-Priv}. \textit{Phase boundaries}: \textit{1}~-- saturation, \textit{2}~-- boiling, \textit{3}~-- deconfinement, \textit{4}~-- hadronization, \textit{CP}~-- critical points. (b): Gas-liquid and plasma phase transitions (GLPT vs. PPT) in hydrogen (figure from~\cite{PPT-Jap-98}). \textit{Phase boundaries} (left to right): GLPT -- saturated vapor, boiling liquid, freezing, melting; PPT -- pressure ionization; CP -- critical points of GLPT and PPT}
\end{figure}	

It should be stressed~\cite{Ios-Hirsch-13} that remarkably similar structure of boundaries for two phase transitions are well known in electromagnetic plasmas. For example it is gas-liquid (left) and ionization-driven (right) phase transitions in dense hydrogen (figure~\ref{tr-h.png}) (see Figure in~\cite{PPT-Jap-98}).

\section{Entropic vs. enthalpic phase transitions}

It is almost evident that two gas--liquid-like PTs, from one side, and two ``delocalization-driven'' PTs (QHPT and PPT), from other side, are similar to each other. This similarity in forms of phase boundaries manifests identity in key physical processes, which rule by phase transformations in both systems in spite of great difference in their densities and temperatures. When one compress isothermally ``vapor'' phase (subscript \textit{V}) in case of GLPT, he reaches saturation conditions. At this moment the system ``jumps'' into ``liquid'' phase (subscript \textit{L}) with \textit{decreasing} of enthalpy and increasing of nega-entropy in accordance with equality rule for Gibbs free energy in 1$^\mathrm{st}$-order PT:
\begin{eqnarray}
&& G_V=H_V-TS_V=H_L-TS_L=G_L\;,
\label{eq1}
\\
&& \Delta G=0 \;\;\; \Leftrightarrow \;\;\; \Delta H=H_V-H_L=T(S_V-S_L)\ge 0\;.
\label{eq2}
\end{eqnarray}
Opposite order of enthalpy and entropy change should be stressed for both ``de-localization-driven'' phase transitions (QHPT and PPT) in figures~\ref{tm-glpt} and~\ref{tm-qhpt}. The both systems, molecular hydrogen (M) and hadronic mixture (H), are ensembles of bound clusters, composed from ``elementary'' particles: protons and electrons in the case of hydrogen and u- and d-quarks in the case of QHPT. The both systems reaches correspondingly ``pressure-deconfinement'' or ``pressure-ionization'' conditions under iso--\textit{T} compression and then jump into deconfinement (Q) or plasma (P) phases correspondingly with \textit{increasing} enthalpy and \textit{decreasing} nega-entropy~(\ref{eq3}), which is just opposite to that in enthalpic PT~(\ref{eq2}):
\begin{equation}
\Delta G_{PPT}=0 \;\;\; \Leftrightarrow  \;\; \Delta H = H_P - H_M = T(S_P - S_M) \ge 0\;,
\label{eq3}
\end{equation}
$$
\Delta G_{QHPT} = 0   \Leftrightarrow   \Delta H = H_Q - H_H = T(S_Q - S_H) \ge 0\;. \eqno(3^*)
$$
Here indexes ``M'' vs. ``P'' and ``H'' vs. ``Q'' denote ``bound'' and ``non-bound'' phases: molecular vs. plasma, and hadron vs. quark phases correspondingly. It is well-known that quark-gluon plasma (QGP) has ``much greater number for degrees of freedom'' than hadronic phase (see e.g.~\cite{Fort-09}). It just means much higher entropy of QGP vs. hadronic phase in thermodynamic terms. This opposite order of enthalpy and entropy changes in two discussed above phase transformation (GLPT and QHPT) is main reason for phase transition classification and terminology accepted and proposed in present paper:
namely \textit{enthalpic} (GLPT) vs. \textit{entropic} (QHPT and PPT) phase transitions.

It is evident that besides well-known ionization-driven (plasma) PT, there are many other candidates for being members of entropy transitions class, namely those PTs, where \textit{delocalization} of bound complexes is just the ruling mechanism for those phase transformations. It is e.g. well-known \textit{dissociation-driven} PT in dense hydrogen, nitrogen and other molecular gases (e.g.~\cite{MPC-10, LHR-10} etc.). It is e.g. more exotic \textit{polimerization}- and \textit{depolimerization-driven} PTs in dense nitrogen and possibly other molecular gases (e.g.~\cite{Yakub-94, Yakub-93-01, Ross-06, BoBo-09, N2-JETP} \textit{etc}.). In all these cases basic feature of entropic~(\ref{eq3}) and enthalpic~(\ref{eq2}) PTs (e.g.~\cite{MOKFF-01}) leads immediately to opposite sign of $P(T)$--dependence at phase coexistence curve in accordance with Clausius -- Clapeiron relation. Hence the positive or negative slope of pressure-temperature phase boundary -- $P(T)_{binodal}$ is the key feature for delimiting of both types of PTs, i.e. enthalpic vs. entropic:
\begin{equation}
\Delta H = T\Delta S  > 0   \Rightarrow   \left(\frac{dP}{dT}\right)_{binodal} > 0\qquad\mbox{(enthalpic PT)},
\label{eq4}
\end{equation}
\begin{equation}
\Delta H = T\Delta S  < 0   \Rightarrow   \left(\frac{dP}{dT}\right)_{binodal} < 0\qquad\mbox{(entropic PT)}.
\label{eq5}
\end{equation}

\subsection{Comparison of entropic vs. enthalpic phase transitions in pressure--temperature plane}
\label{sec:TestSubsection}

Exponentially increasing (VdW-like) form of \textit{P--T} phase diagram for ordinary GLPT in hydrogen and other substances is well-known. Similar \textit{P--T} dependence of GLPT in nuclear matter was calculated many times, e.g.~\cite{TRKBW-10, SDM-09, HDSI-13} \textit{etc.} (see Fig. 3 left). In contrast to that \textit{P--T} phase diagram of QHPT (Fig. 3 right) is known, but not widely known (\cite{SaMi-10, Rand-12, Rand-CPOD}). It was calculated recently in~\cite{HDSI-13}. Both phase transitions, GLPT and QHPT, have \textit{opposite} $P(T)$ behavior in agreement with~(\ref{eq4}) and~(\ref{eq5}). This fact is not recognized properly as a general phenomenon~\cite{ILI-S-H-PT-13, NUFRA-13}.

\begin{figure}[t]
\subfigure[Gas-liquid phase transition]{
\includegraphics[width=0.49\columnwidth]{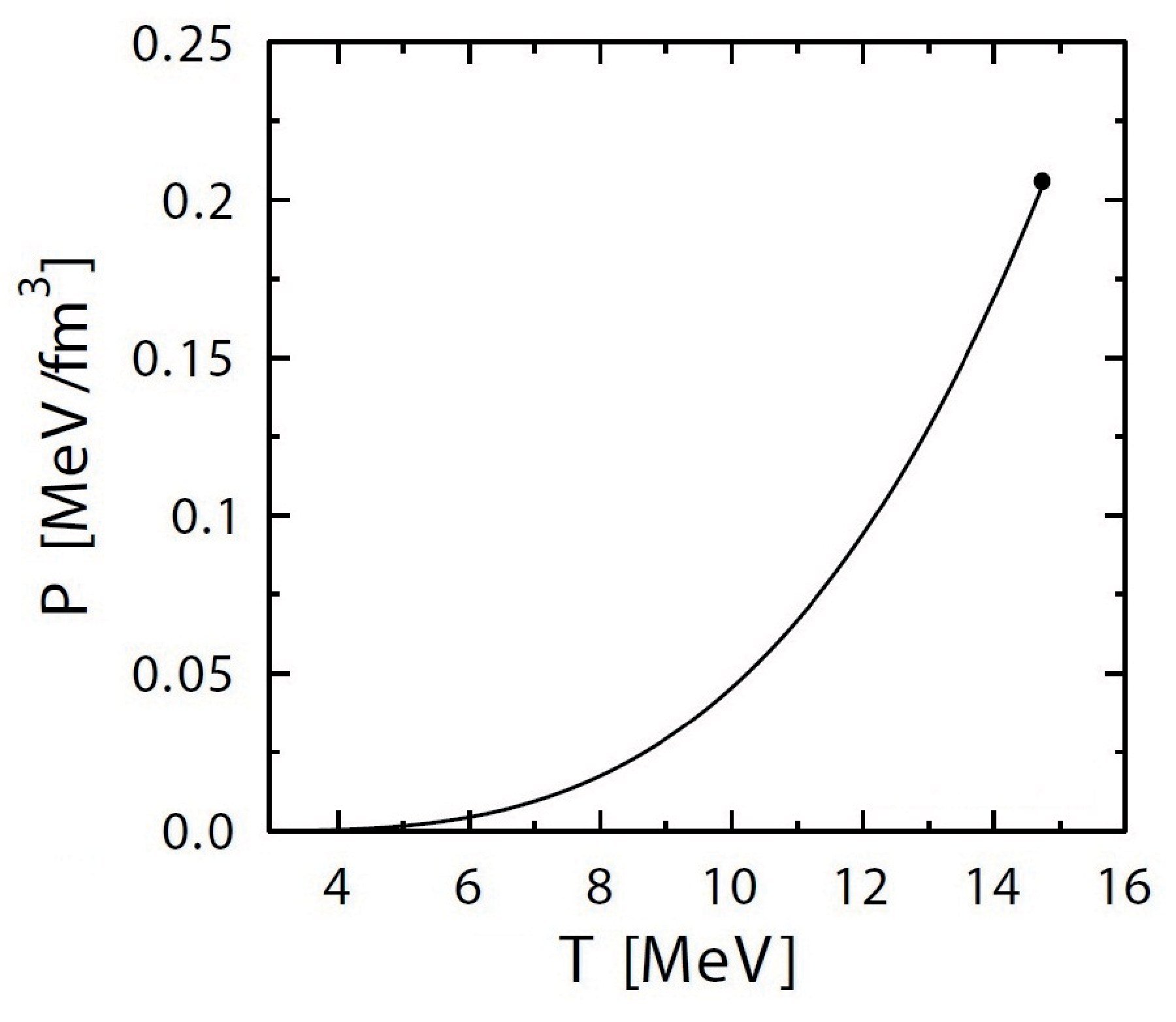}
\label{fig:pt-glpt}}
\hfill
\subfigure[Quark-hadron phase transition]{
\includegraphics[width=0.45\columnwidth]{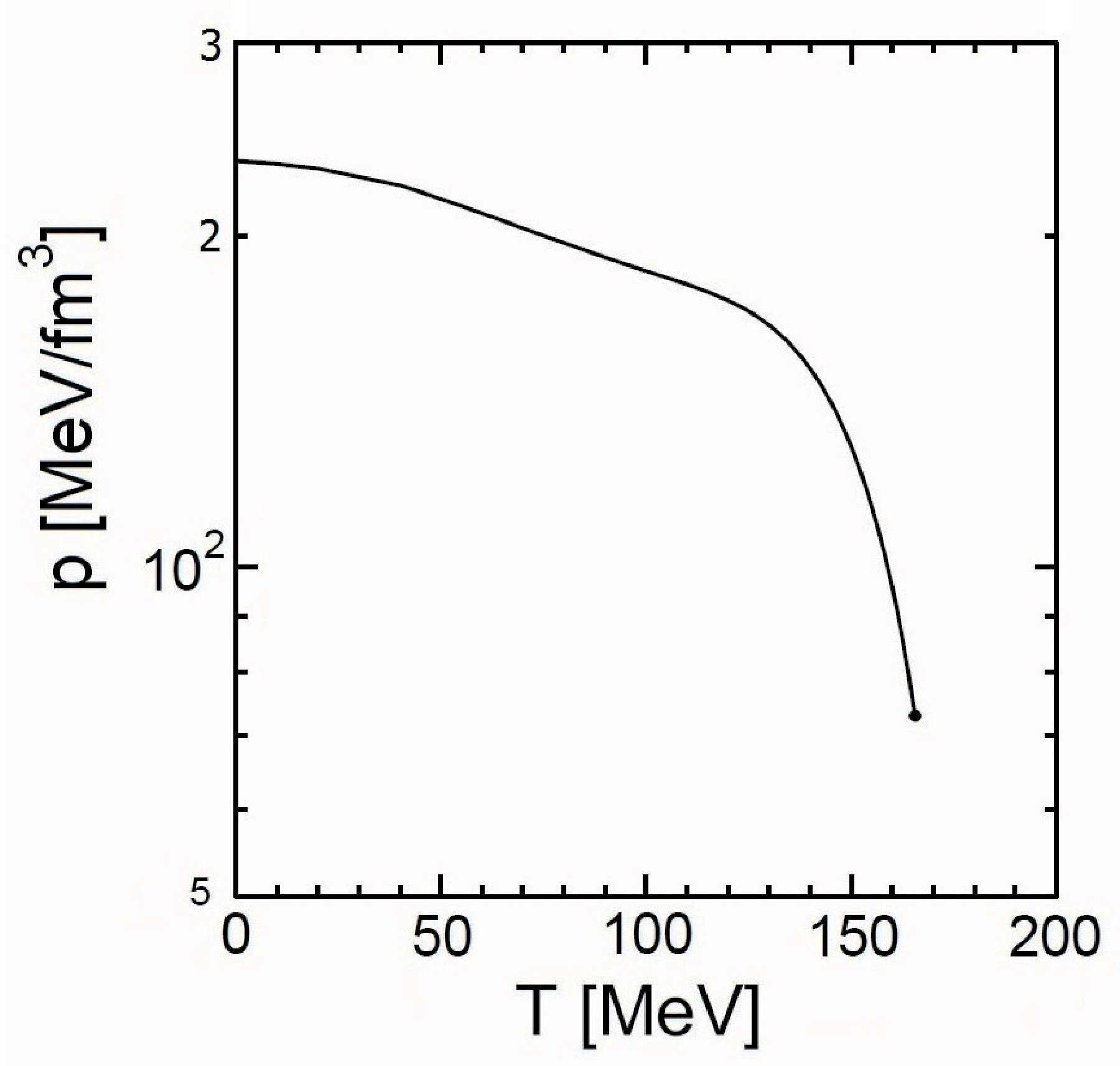}
\label{fig:pt-qhpt}
}
\caption{Pressure--temperature phase diagram of gas-liquid (a) and quark-hadron (b) phase transition in symmetric nuclear matter (figs from~\cite{HDSI-13}).}
\end{figure}

\subsection{Enthalpic \textit{vs.} entropic phase transitions in pressure--density (pressure--specific volume) plane}

The most striking difference between entropic vs. enthalpic types of phase transitions could be demonstrated in comparison of their \textit{P--V} phase diagrams. It should be noted that just this phase diagram is the most important for analysis of many dynamic processes in dense plasma: e.g. shock or isentropic compression as well as adiabatic expansion, including anomalous shock rarefaction waves. \textit{P--V} phase diagram for VdW-like GLPT in ordinary substances is well known. GLPT in symmetric Coulombless nuclear matter has the same structure (see e.g.~\cite{TRKBW-10, SDM-09} etc.). In contrast to that the \textit{P--V} phase diagram for phase transitions, which are claimed as \textit{entropic} PTs (ionization-, dissociation-, polymerization- and depolymerization-driven PTs and more general -- ``delocalization-driven'' PTs~\cite{ILI-S-H-PT-13}) were not explored properly yet. In particular, the \textit{P--V} phase diagram for Quark-Hadron phase transition (QHPT) was not explored up to date. It is just in process on the base of QHPT calculations in~\cite{HDSI-13}.

\begin{SCfigure}[][ht!]
\includegraphics[width=0.6\textwidth]{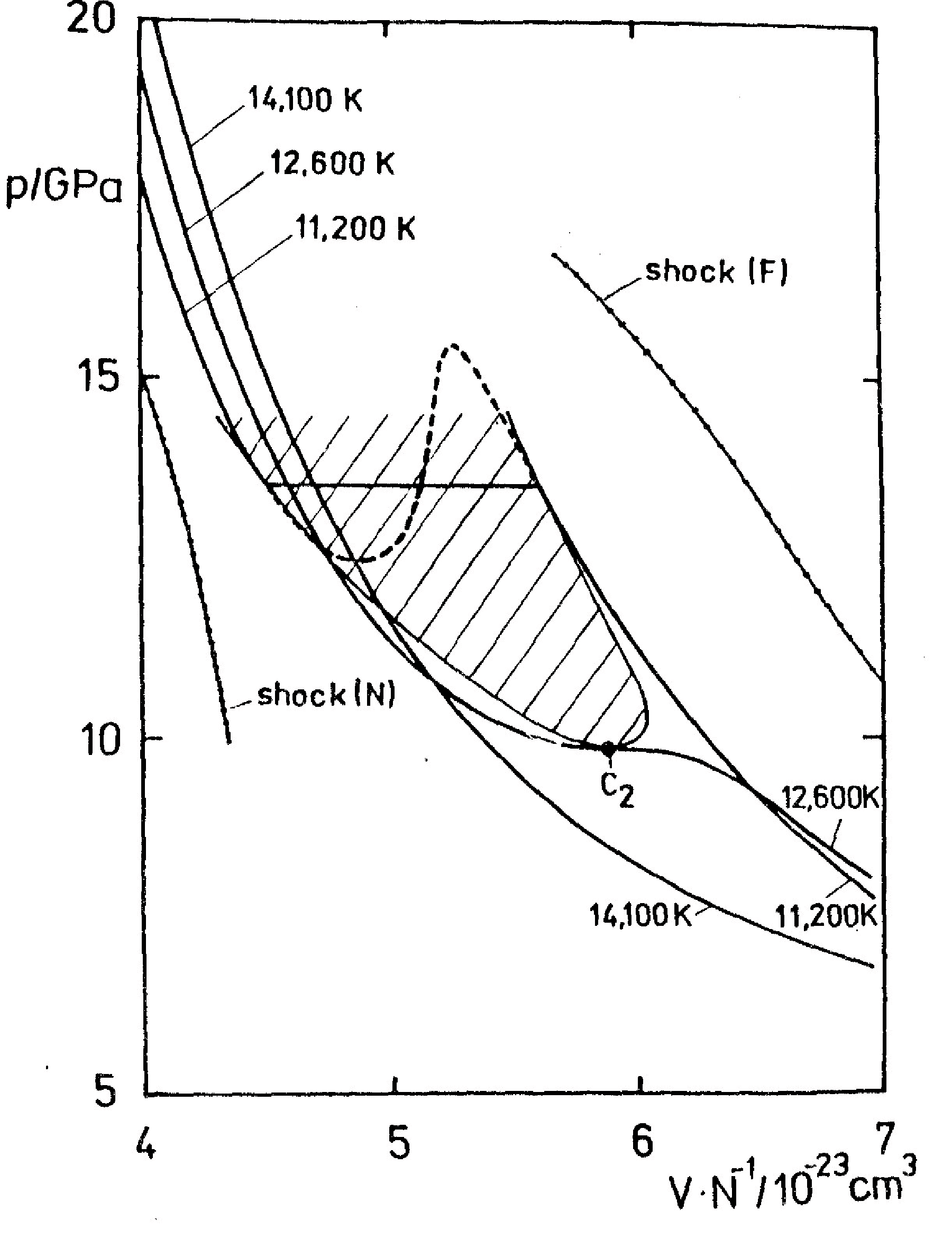}
\caption{\textit{P--V} phase diagram of hypothetical ionization-driven (``plasma'') phase transition in xenon. Solid lines -- calculated isotherms for $T \le T_c$ and $T \ge T_c$ ($T_c\approx 12600$ K). Initial VdW-loop and equilibrium part of two-phase isotherm are shown for $T = 11 200$ K. Shaded area -- two-phase region. $C_2$ -- critical point. Dot-solid lines -- estimated parameters of shock adiabats; N -- Nellus \textit{et al.}, F -- Fortov \textit{et al}. (Figure from~\cite{DDR-80}).}
\label{fig:pr-xe}
\end{SCfigure}

A good example of typical $P-V$ phase diagram for entropic ionization-driven (``plasma'') phase transition (PPT) in xenon is exposed at fig.~\ref{fig:pr-xe} accepted from~\cite{DDR-80} (see also fig.~III.6.11a in~\cite{IoSta-00}). Even more clearly anomalous thermodynamics in the vicinity of two-phase region for entropic PTs is illustrated at fig.~\ref{fig:pr-gvk} (below) for example of \textit{dissociation-driven} phase transition in simplified EOS (SAHA-model) for shock-compressed deuterium
(see also fig.~4 in~\cite{DDR-80}).

\section{Abnormal thermodynamics in the vicinity of two-phase region of entropic PTs}
\label{sec:AnomalousThermodynamicsInTheVicinityOfTwoPhaseRegionOfEntropicPTs}

Several important features of \textit{abnormal thermodynamic behavior} in two-phase region of this PPT and in its close vicinity
are demonstrated at the fig.~7 from~\cite{DDR-80} and fig.~8 from~\cite{GrIo-09}:
\begin{enumerate}
\item -- more than one isotherms come through the critical point of entropic PT in $P$--$V$ plane projection 
(see e.g. fig.~\ref{fig:pr-gvk});
\item -- several isotherms below and above critical isotherm \textit{cross each other} not only in two-phase region (it is obligatory for entropic PT) but in its close one-phase vicinity;
\item -- many low-$T$ isotherms $P(V)_T$ and $V(P)_T$ lay \textit{above}, at least partially, of high-$T$ isotherms (i.e. at higher $P$ and higher $V$, correspondingly).
\end{enumerate}

\textit{\underline{Comment}}:
It should be stressed that features (2) and (3) above means abnormal \textit{negative sign} of thermal pressure and thermal expansion coefficients in discussed area around and within the two-phase region of entropic PT. It means that $(\partial P/\partial T)_V < 0$ and $(\partial V/\partial T)_P < 0$. It is reasonable to assume that the violation of (ii) and (iii) occurs in \textit{isolated} $P$--$V$ \textit{area}, which is located between the regions with normal thermodynamics \{i.e. with positive sign of $(\partial P/\partial T)_V$ and $(\partial V/\partial T)_P$ \}.

\subsection{Abnormal negativity of cross second derivatives}
\label{sec:AnomalousNegativenessOfCrossSecondDerivatives}

Negativity of two second derivatives $(\partial P/\partial T)_V$ and $(\partial V/\partial T)_P$ is never isolated event. In frames of straightforward thermodynamic technique it proves to be equivalent to simultaneous negativity for infinite number of other accompanying second derivatives for thermodynamic potentials. In particular, negativity of $(\partial P/\partial T)_V$ and $(\partial V/\partial T)_P$ combined with the positivity of heat capacities $C_V$ and $C_P$ (as obligatory conditions of thermodynamic stability) leads to the negativity for following six cross derivatives: (here $U$, $S$ and $H$ are internal energy, entropy and enthalpy):
%

\begin{align}
(\partial P/&\partial T)_V   &&  \leftrightarrow  &   (\partial P/&\partial S)_V  &&  \leftrightarrow   &  (\partial P/&\partial U)_V  \label{dd-V}\\
\updownarrow&                &&                   &   \updownarrow&               &&                    &  \updownarrow&                     \notag\\
(\partial V/&\partial T)_P   &&  \leftrightarrow  &   (\partial V/&\partial S)_P  &&  \leftrightarrow   &  (\partial V/&\partial H)_P  \label{dd-P}\\
            &                &&  \updownarrow     &                               &&  \updownarrow      &              &                     \notag\\
            &                &   (\partial S&/\partial V)_T  &  \leftrightarrow  && (-\partial S/&\partial P)_T   \label{dd-T}\\	
            &                &&  \updownarrow     &                               &&  \updownarrow      &              &                     \notag\\					
            &                &   (\partial T&/\partial P)_S  &  \leftrightarrow  && (-\partial T/&\partial V)_S   \label{dd-S}			
\end{align}

\begin{enumerate}
\item It should be stressed and clarified that all ten cross derivatives in~(\ref{dd-V}),~(\ref{dd-P}),~(\ref{dd-T}),~(\ref{dd-S})  
are positive or negative or \textit{equal to zero} simultaneously.
\item Note that three cross derivatives in~(\ref{dd-V}) and~(\ref{dd-P}) are equivalent to three conventional thermodynamic coefficients:
\begin{enumerate}
\item -- thermodynamic Gruneizen parameter,  $Gr \equiv V(\partial P/\partial U)_V $,
\item -- thermal expansion coefficient, $\alpha_T \equiv V^{-1}(\partial V/\partial T)_P $,
\item -- isochoric thermal pressure coefficient,  $\alpha_P \equiv P^{-1}(\partial P/\partial T)_V$,
\end{enumerate}

\textit{\underline{Comment}}:
Simultaneous  positivity or negativity of two cross derivatives: Gruneizen parameter ($Gr$) and thermal expansion coefficient ($\alpha_T$) is well known and, for example, was used for explanation of abnormal properties of shock compression of nitrogen (see e.g.~\cite{Yakub-94, Yakub-93-01, Nellis-1991, N2-JETP}) and anomalies in shock compression of silica (see~\cite{Medved-UFN} and discussion in~\cite{Bra-12})

%
%

\textit{\underline{Comment}}:
One should be careful with the sign of two above written derivatives within the two-phase region of (congruent) entropic phase transition, first one in~(\ref{dd-P}) and last one in~(\ref{dd-T}): i.e. $(\partial V/\partial T)_P$ and $(- \partial S/\partial P)_T$. Both the derivatives tend to infinity (!) within the two-phase region, where isotherms and isobars coincide. But the ``sign'' of this infinity is conjugated with the sign of all eight other finite derivatives in~(\ref{dd-V}),~(\ref{dd-P}),~(\ref{dd-T}) and~(\ref{dd-S}). It means that the both derivatives, $(\partial V/\partial T)_P$ and $(- \partial S/\partial P)_T$, tend to minus infinity in the case of negative (anomalous) sign of all other finite derivatives in~(\ref{dd-V}),~(\ref{dd-P}),~(\ref{dd-T}) and~(\ref{dd-S}):
\begin{equation}
(\partial V/\partial T)_P\rightarrow-\infty\Leftrightarrow(-\partial S/\partial P)_T\rightarrow-\infty\qquad\text{(in two-phase region)}.
\label{inf}
\end{equation}
\end{enumerate}
\textit{\underline{Comment}}: Negativity of all notified above second cross derivatives leads to important consequences in mutual order and behavior of all thermodynamic isolines in $P$--$V$ plane, i.e. isotherms, isentropes and shock adiabats first of all.
\begin{enumerate}
\item -- Negativity of $(\partial P/\partial T)_V$ leads to abnormal crossing and interweaving of isotherms;
\item -- Negativity of $(\partial P/\partial S)_V$ leads to abnormal crossing and interweaving of isoentropies;
\item -- It leads also to abnormal relative order of  shock adiabats vs. isoentropies and isotherms;
\item -- Negativity of $(\partial P/\partial U)_V$ leads to abnormal relative order and crossing of shock (Hugoniots) adiabats.
\end{enumerate}

It is known that anomalous crossing of Hugoniots adiabats follows from negativity of Gruneizen coefficient 
(see e.g.~\cite{Medved-UFN}). 
So-called ``shock cooling'' of nitrogen in reflected shocks~\cite{Nellis-1991} could also be explained with assumption of negative Gruneizen coefficient (see e.g.~\cite{Yakub-94, Yakub-93-01, Ross-06}). Thus abnormal negativity of whole group of cross derivatives~(\ref{dd-V}),~(\ref{dd-P}),~(\ref{dd-T}),~(\ref{dd-S}) is of primary importance for the \textit{hydrodynamics} of adiabatic flows, e.g. shock compression, isentropic expansion, adiabatic expansion into vacuum, spinodal decomposition etc. All these problems should be discussed separately~\cite{ZeRa-08, LaLi-86} (see also~\cite{Ios-Nega-Gr}).

\section{Abnormal structure of iso-lines in $P$--$V$ plane and additional metastable section within two-phase region of entropic phase transition}
\label{sec:AnomalousBehaviorOfIsoLinesInPVPlaneAndNewAdditionalRegionOfMetastableStateWithinTwoPhaseRegionOfEntropicPT}

\begin{figure}[t!]
\centering
\includegraphics[width=0.6\textwidth]{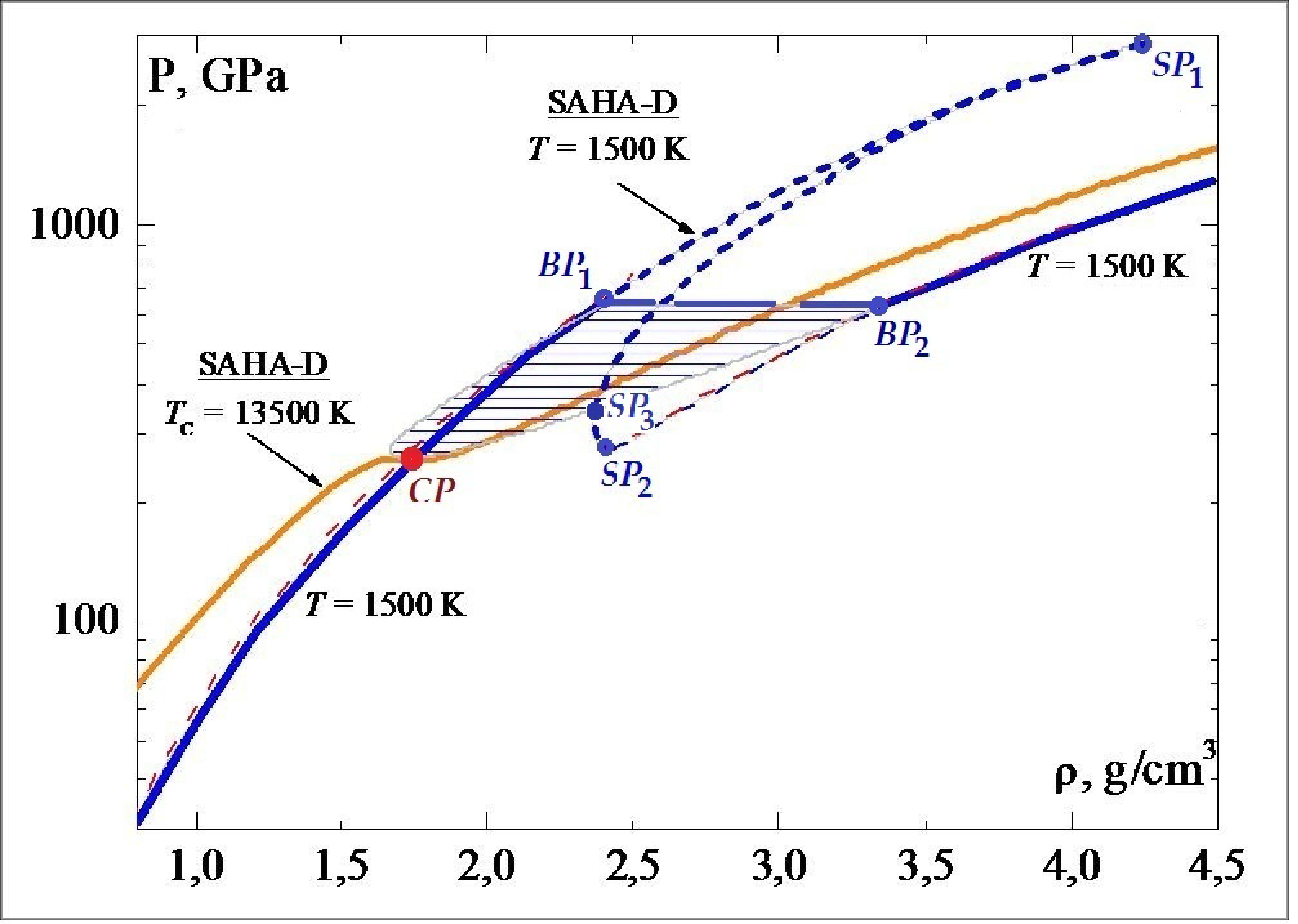}
\caption{$P$--$\rho$ phase diagram for hypothetical dissociation-driven (entropic) phase transition in dense deuterium (SAHA-D code~\cite{GrIo-09}). \textit{Solid lines} -- calculated critical ($T_c \approx 13500$ K) and subcritical ($T << T_c$) isotherms. Dashed curves at $T = 1500$ K -- initial meta- and unstable parts of isotherm. {\em Shaded area} -- two-phase region; \textit{CP} -- critical point; \textit{BP$_1$, SP$_1$, BP$_2$} and \textit{{SP}$_2$} -- binodal and spinodal points for rare (1) and dense (2) phases. \textit{SP}$_3$ -- new additional spinodal
(\ref{SP-N}).
Dashed curves at $T$ = 1500 K -- \textit{two} unstable and \textit{three} metastable parts of initial isotherm. \textit{SP}$_1$--\textit{SP}$_3$ -- new metastable part of $S$-PT (Figure from \cite{GrIo-09}).}
\label{fig:pr-gvk}
\end{figure}

One meets anomalous behavior of isotherms within and near the two-phase region of discussed entropic phase transition at sufficiently low temperature, namely:

\begin{enumerate}
\item\label{5.1.1} -- appearance of \textit{return-point} behavior of metastable part of isotherm in upper spinodal region at low-density branch of isotherm (see e.g. upper end-point at $T = 1500$ K at fig.~\ref{fig:pr-gvk});
\item\label{5.1.2} -- one more \textit{third metastable section} with positive compressibility (i.e. $(\partial P/\partial V)_T < 0$) appears within two-phase region of entropic transition in contrast to conventional structure of metastable and unstable parts in enthalpic Van-der-Waals-like (gas-liquid) phase transition. This new metastable section lays \textit{between two unstable} parts of low enough subcritical isotherms within spinodal region of isotherm (see e.g. $T = 1500$ K at fig.~\ref{fig:pr-gvk}). Features~(\ref{sec:AnomalousBehaviorOfIsoLinesInPVPlaneAndNewAdditionalRegionOfMetastableStateWithinTwoPhaseRegionOfEntropicPT}.\ref{5.1.1}) and~(\ref{sec:AnomalousBehaviorOfIsoLinesInPVPlaneAndNewAdditionalRegionOfMetastableStateWithinTwoPhaseRegionOfEntropicPT}.\ref{5.1.2}) are in contrast to standard behavior of gas-liquid PT, where one unstable part of isotherm divides two metastable parts in ordinary VdW-loop;	
\item\label{5.1.3} -- one more \textit{new spinodal} (i.e. boundary between metastable and unstable parts within two-phase region) appears, which bounds this \textit{third metastable section}. It is the locus of points obeying condition~(\ref{SP-N}), which is \textit{opposite} to well-known condition of standard spinodal for conventional (enthalpic) gas-liquid phase transition~(\ref{SP-O}):

\begin{equation}
\text{Conventional spinodal (enthalpic PT)}\qquad(\partial P/\partial V)_T = 0,
\label{SP-O}
\end{equation}
\begin{equation}
\text{New additional spinodal (entropic PT)}\qquad(\partial P/\partial V)_T = \infty.
\label{SP-N}
\end{equation}

\item\label{5.1.4} -- in addition to conventional critical point \{i.e. the point, where $(\partial P/\partial V)_T = 0$ and $(\partial^2 P/\partial V^2)_T = 0$\}, which is ``upper'' in $T$--$V$ plane, and is ``lower'' in $P$--$V$ plane, one more \textit{new singular point} (notation below -- \textit{NSP}) appears within two-phase region of entropic PT at low enough subcritical isotherm. Isothermal compressibility is equal to zero in this NSP \{i.e. $(\partial P/\partial V)_T = \infty$\} in contrast to the ordinary critical point, where isothermal compressibility is infinite, i.e. $(\partial V/\partial P)_T = \infty$! This new singular point obeys to~(\ref{SP-N}) and closes discussed third metastable section of entropic PT from above in $T$--$V$ plane (upper end-point) and from below in $P$--$V$ plane (lower end-point). More detailed discussions and illustrations of all mentioned above new objects are in progress~\cite{Ios-Nega-Gr}.
\newcounter{enumTemp}
\setcounter{enumTemp}{\arabic{enumi}}
\end{enumerate}

Next anomalous features, exposed at fig.~\ref{fig:pr-xe} and~\ref{fig:pr-gvk}, should be emphasized in addition to those mentioned above:
\begin{enumerate}
\setcounter{enumi}{\theenumTemp}
\item\label{5.1.5} -- spinodal cupola, which is \textit{always} located \textit{inside} binodal cupola in the case of enthalpic VdW-like PT, now located \textit{partially outside} of binodal area in the case of entropic PT (fig.~\ref{fig:pr-gvk})
\item\label{5.1.6} -- spinodal point of rare branch of isotherm (it resembles ``overcooled vapor'' in VdW phase transition) may have \textit{higher density} than spinodal point of dense branch of isotherm (which resembles ``overheated liquid'') at low enough subcritical temperature (fig.~\ref{fig:pr-gvk}).
\end{enumerate}

\section{``Zero-Boundary'' -- the border for the region of abnormal thermodynamics}
\label{sec:ZeroLineTheBoundaryForTheAbnormalThermodynamicsRegion}

The region of discussed above abnormal thermodynamics (AT-region) always accompanies entropic phase transitions.
At the same time it could be \textit{isolated region} without any $\mathrm{1^{st}}$-order phase transition-like discontinuity.
\textit{Two variants} of the boundary of such AT-region should be distinguished.
\begin{enumerate}
\item
AT-region coincides with two-phase region of entropic PT so that AT-region and PT-region have common boundary.
This is just the case for so-called \textit{non-isostructural} phase transitions, i.e. for transitions between phases with principally different structures, where coexisting phases could not be transformed continuously one into another.
Well-known examples of non-isostructural PTs are crystal--fluid PT (melting) and polymorphic PTs between phases with different crystalline lattices (e.g. \textit{bcc-fcc etc}.).
Great number of examples for such PTs with (at least partially) decreasing $P(T)$-dependence are well known (see e.g. ``Generalized phase diagram'' at fig. 16 in \cite{Bra-97}).
\item
The boundary of AT-region (at least partially) located in the region of regular thermodynamics with \textit{continuous} transition from area with normal thermodynamics (positive sign of all cross derivatives~(\ref{dd-V}),~(\ref{dd-P}),~(\ref{dd-T}) and~(\ref{dd-S})) to area with abnormal thermodynamics (negative sign of these derivatives).
It is so in particular for the case of isolated AT-region (without phase transition at all) and in the case of \textit{isostructural} PT, like all fluid-fluid phase transitions, where critical point (or even several critical points!) must exist.
New thermodynamic object appears in this latter case -- the locus of points where all cross derivatives~(\ref{dd-V}),~(\ref{dd-P}),~(\ref{dd-T}) and~(\ref{dd-S}) are \textit{equal to zero simultaneously}.
(Having no widely accepted title for this object we would use below the notation ``Zero-Boundary'' -- ZB).
Remarkable features in behavior of thermodynamic iso-lines upon and in the vicinity of ZB, as well as main consequences for zero-value of cross derivatives~(\ref{dd-V}),~(\ref{dd-P}),~(\ref{dd-T}),~(\ref{dd-S}) for main hydrodynamic properties of adiabatic processes, e.g. shock and isentropic compression and expansion etc., will be later discussed separately.
\item
The most evident thermodynamic properties at any point of Zero-Boundary are following:
\begin{enumerate}
	\item
	Isobaric and isochoric heat capacities are equal to each other:
	\begin{equation}
	C_p \equiv (\partial H/\partial T)_P  =  (\partial U/\partial T)_V  \equiv C_V.
\label{ZL-1}
	\end{equation}
	\item
	Isothermal and isentropic speeds of sound are equal to each other:
	\begin{equation}
	a_S \equiv (\partial P/\partial\rho)^{1/2}_S  =  (\partial P/\partial\rho)^{1/2}_T  \equiv a_T\label{ZL-2}
	\end{equation}
	\item
	The slopes of four iso-lines: e.g. iso-$T$, iso-$S$, iso-$U$, iso-$H$ (temperature, entropy, internal energy and enthalpy) and slope of shock adiabat (Hugoniot) in $P$--$V$-plane are equal to each other at Zero-boundary:
\begin{equation}
(\partial P/\partial V)_T=(\partial P/\partial V)_S=(\partial P/\partial V)_U=(\partial P/\partial V)_H=(\partial P/\partial V)_{Hug}.
\label{ZL-3}
\end{equation}
\end{enumerate}

\end{enumerate}


\section{Multilayered structure of thermodynamic surfaces $X(P,V)$ in the region of abnormal thermodynamics}
\label{sec:MultilayeredStructureOfThermodynamicXPVSurfacesInTheRegionOfAnomalousThermodynamics}

All mentioned above anomalies have clear geometric interpretation: -- temperature, entropy and internal energy surfaces as a functions of pressure and density, e.g. $T(P,V)$, $S(P,V)$ and $U(P,V)$, have \textit{multi-layered} structure over the region of anomalous thermodynamics in $P$--$V$ plane in the case of \textit{all} entropic phase transitions. It is valid in particular, for all discussed above ``delocalization-driven'' phase transitions like ionization-, dissociation-, depolymerization-driven PTs, as well as for quark-hadron transition (QHPT) \textit{etc}. Again, one should distinguish two variant of such multi-layered structure of $T$, $S$ and $U$ over $P$--$V$ plane: (\textit{i}) -- when AT-region coincides with two-phase region of a phase transition, so that both have the same common boundary, which is the locus of break in $T$, $S$, $U$-surfaces; and (\textit{ii}) -- when AT-region restricted (at least partially) by the separate boundary (outside of the two-phase boundary itself) with \textit{zero value} for all cross derivatives mentioned in eq-s~(\ref{dd-V}),~(\ref{dd-P}),~(\ref{dd-T}) and~(\ref{dd-S}) (i.e. ``Zero-boundary'').

\section{What should one classify in case of unexplored phase transition}
\label{sec:WhatShouldWeClassifyInCaseOfUnexploredPhaseTransition}

Keeping in mind discussed above difference between enthalpic and entropic phase transitions we ought to summarize main features,
which should be classified, when one meets unexplored phase transition (see e.g.~\cite{IoSta-00, Ios-Acta}):
\begin{itemize}
\item
Is this PT of $\mathrm{1^{st}}$ or $2^\mathrm{nd}$-order?
\item
Is this PT enthalpic or entropic? (this paper)
\item
Is this PT isostructural or non-isostructural (like, for example, gas-liquid PT \textit{vs}. crystal-fluid PT)?
\item
Is this PT congruent or non-congruent (see e.g.~\cite{IGYRF-03, Ios-Acta, HDSI-13})?
\item
Do we use Coulombless approximation in description of this PT (see e.g.~\cite{HDSI-13}), or we take into account all consequences of long-range nature of Coulomb interaction?
\item
What is the scenario of \textit{equilibrium} phase transformation in two-phase region? Is it macro- or mesoscopic one (e.g. ``structured mixed phase (SMP)'' scenario (or so-called ``pasta-plasma''))?
\end{itemize}

In this paper we made accent on analysis of following combination: $\mathrm{1^{st}}$-order, entropic, congruent, isostructural and/or non-isostructural, Coulomb-included phase transitions without possibility of SMP (pasta) scenario.
Phase transitions with other more complicated combination of basic features would be discussed in following papers.

\section*{Conclusions}

\begin{itemize}

\item
Widely accepted visible equivalence of gas-liquid-like and quark-hadron (deconfinement) phase transitions in high energy density nuclear matter is illusive.
\item
Both phase transitions belong to fundamentally different sub-classes: gas-liquid PT is enthalpic one, while quark-hadron PT is entropic one.
\item
Properties of entropic and enthalpic PTs differ significantly from each other.
\item
Pressure-temperature dependence of phase boundary for enthalpic phase transition ($H$-PT) and entropic one ($S$-PT) have different sign i.e. $(dP/dT)_{H-PT} \ge 0$ \textit{vs.} $(dP/dT)_{S-PT} \le 0$.
\item
Isotherms of entropic PT have anomalous behavior within the two-phase region at sufficiently low subcritical temperature. There is abnormal order for sequence of stable, metastable and unstable parts: e.g. stable-I / metastable-I / unstable-I / \textit{metastable}-III / unstable-II / metastable-II / stable-II.
\item
Binodals and spinodals of entropic PT have anomalous order in $P$--$V$ plane. Isothermal spinodal [e.g. $(\partial P/\partial V)_T = 0$] may be located \textit{outside} binodal of entropic PT at low enough subcritical temperature.
\item
Two-phase region and its close vicinity for entropic PT obey to abnormal thermodynamics. Namely: negative Gruneizen parameter, negative thermal and entropic pressure coefficients, negative thermal expansion coefficient \textit{etc}. Besides there are anomalous order and intersections of isotherms, isentropes and abnormal order and intersections of shock adiabats (Hugoniots) \textit{etc}.
\item
All the features of discussed entropic phase transitions and accompanying abnormal thermodynamics region have transparent geometrical  interpretation -- multi-layered structure of thermodynamic surfaces for temperature, entropy and internal energy as a pressure--volume functions, e.g. $T(P,V)$, $S(P,V)$ and $U(P,V)$.
\item
Deconfinement-driven (QHPT) and ionization-driven ``plasma'' phase transitions (PPT) as well as dissociation- and depolymerization-driven PTs in $\mathrm{N_2}$ \textit{etc}. are entropic PTs, and hence they have many common features in spite of many order difference in density and temperature.
\item
It is promising to investigate entropic PTs experimentally, for example via volumetric heating by heavy ion beams (HIB), in strong shock compression and subsequent isentropic expansion, in quasi-isobaric expansion with exploding foil technique and via surface laser heating \textit{etc}.
\item
It is especially promising also to investigate entropic PTs theoretically in frames of traditional thermodynamic models (chemical picture) and via \textit{ab initio} approaches.
\end{itemize}

\subsection*{Acknowledgement}

We express our thanks to the organizers of the CSQCD IV conference for providing an excellent atmosphere which was the basis for inspiring discussions with all participants. We have greatly benefitted from this.
Author acknowledges V. Gryaznov for collaboration and for skilful calculations in obtaining phase diagram of modeling dissociative phase transition in deuterium, and M. Hempell and V. Dexheimer for collaboration and permission to use joint density-temperature phase diagram
for GLPT and QHPT. Author acknowledges also V. Fortov, D. Blaschke and J. Randrup for helpful and fruitful discussions.
This work was supported by the Presidium RAS Scientific Program ``Physics of extreme states of matter''

\end{document}

%% file: econfmacros-csqcd4.tex
\def\beq{\begin{equation}}
\def\eeq#1{\label{#1}\end{equation}}
\def\eeqn{\end{equation}}

\def\beqa{\begin{eqnarray}}
\def\eeqa#1{\label{#1}\end{eqnarray}}
\def\eeqan{\end{eqnarray}}

\let\bar=\overbar

\def\Dslash{\not{\hbox{\kern-4pt $D$}}}
\def\dslash{\not{\hbox{\kern-2pt $\del$}}}

\def\msb{{\bar{\ssstyle M \kern -1pt S}}}

\usepackage{fancyhdr,graphicx}